\documentclass[aps,prd,reprint,nofootinbib,longbibliography]{revtex4-1}

\usepackage{amsmath}
\usepackage{mathtools}
\usepackage{graphicx}
\usepackage[normalem]{ulem}
\usepackage[com pat=1.1.0]{tikz-feynman}
\usepackage{tikz}
\usepackage{aas_macros}
\usetikzlibrary{external}
\usepackage{comment}
\usetikzlibrary{arrows}
\usepackage[colorlinks=true, allcolors=blue]{hyperref}

\newcommand\barparena[1]{\overset{%
   \scriptscriptstyle(-)}{#1}}

\begin{document}

\title{General-Relativistic Quantum-Kinetics Neutrino Transport}

\author{Hiroki Nagakura}
\email{hiroki.nagakura@nao.ac.jp}
\affiliation{Division of Science, National Astronomical Observatory of Japan, 2-21-1 Osawa, Mitaka, Tokyo 181-8588, Japan}

\begin{abstract}
We developed a new general-relativistic quantum-kinetics neutrino transport code, GRQKNT, for numerical studies of quantum kinetics of non-equilibrium neutrinos in six-dimensional phase space. This code is intended for use in both local and global simulations of neutrino transport in core-collapse supernova and binary neutron star merger. It has been widely recognized that global simulations of collective neutrino oscillations, in particular fast neutrino-flavor conversions, require unfeasible computational resources due to large disparity of scales between flavor conversion and astrophysical phenomena. We propose a novel approach to tackle the issue. This paper is devoted to describe the philosophy, design, and numerical implementation of GRQKNT with a number of tests ensuring correct implementation of each module. The code is based on a discrete-ordinate Sn method, finite-difference realization of mean-field quantum kinetic equation. The transport equation is solved based on a conservative formalism, and we use a fifth-order weighted essentially non-oscillatory scheme with fourth-order TVD Runge-Kutta time-integration. The transport module is designed to work with arbitrary spacetimes and currently three different stationary spacetimes (flat spacetime, Schwarzschild black hole, and Kerr black hole) are implemented. The collision term including neutrino emission, absorption, and momentum-exchanged scatterings are also implemented into our code. The oscillation Hamiltonian consists of vacuum, matter, and self-interactions. Both two- and three neutrino-flavor scenarios can be applied. Fluid-velocity dependences in transport-, collision-, and oscillation modules, are also treated self-consistently by using two-energy-grid technique, that has been already established in our another code with full Boltzmann neutrino transport.
\end{abstract}
\maketitle

\section{Introduction}\label{sec:intro}
In hot and dense medium arising in core-collapse supernova (CCSN) and binary neutron star merger (BNSM), neutrinos play a key role in transporting energy, momentum, and lepton-number. Once neutrinos are produced by weak interactions, they travel in medium. A fraction of these neutrinos experiences scatterings with or reabsorption onto matter, that converts neutrino energy and momentum into those of matter, and then affects the fluid-dynamics. The neutrino emission and absorption can also change the electron-fraction of matter. This has a direct influence on the chemical composition, thus accounting for nucleosynthesis in the ejecta. These things highlight the importance of developing accurate modelling of neutrino radiation field.

Decades of progress on numerical simulations of CCSN and BNSM with Boltzmann neutrino transport or its approximate methods have improved our understanding of rolls of neutrinos in fluid dynamics, nucleosynthesis, and observational consequences such as neutrino signal. The classical treatment of neutrino kinetics, i.e., Boltzmann neutrino transport is justified as long as neutrinos are stuck in flavor eigenstates, which was naturally expected due to large matter potential in CCSN and BNSM environments (see, e.g., \cite{2000PhRvD..62c3007D}). On the other hand, in dense neutrino environments, neutrino-neutrino self-interactions give rise to refractive effects \cite{1992PhLB..287..128P}, indicating that the neutrino dispersion relation is modified. This potentially triggers large neutrino-flavor conversion \cite{2010ARNPS..60..569D}.

It has been suggested that various types of flavor conversions emerge by neutrino self-interactions. For instances, slow neutrino-flavor conversion, that is driven by interplay between vacuum- and self-interactions, leads to synchronized-, bipolar-, and spectral split phenomena (see \cite{2010ARNPS..60..569D} and references therein). Matter neutrino resonances may occur BNSM or collapsar environment \cite{2012PhRvD..86h5015M,2016PhRvD..93d5021M,2016PhRvD..94j5006Z}, in which the dominance of electron-type anti-neutrinos ($\bar{\nu}_{e}$) over electro-type neutrinos ($\nu_{e}$) cancels the matter potential, which induces the similar resonant phenomena as Mikheyev-Smirnov-Wolfenstein (MSW) effect. Collisional instability is a new type of instability of flavor-conversion, in which the disparity of matter interactions between neutrinos- and anti-neutrinos induces the flavor conversion \cite{2021arXiv210411369J}. Fast neutrino-flavor conversion (FFC) has received increased attention from the community, since it would ubiquitously occur in both CCSN \cite{2019PhRvD.100d3004A,2019ApJ...886..139N,2020PhRvR...2a2046M,2020PhRvD.101d3016A,2020PhRvD.101b3018D,2021PhRvD.103f3033A,2020PhRvD.101f3001G,2021PhRvD.103f3013C,2021PhRvD.104h3025N,2022ApJ...924..109H} and BNSM \cite{2017PhRvD..95j3007W,2017PhRvD..96l3015W,2020PhRvD.102j3015G,2021PhRvL.126y1101L,2022arXiv220316559J}. Since the growth rate of FFC is proportional to neutrino number density, it would offer the fastest growing mode of flavor conversion in CCSN and BNSM (see also recent reviews, e.g., \cite{2016NuPhB.908..366C,2020arXiv201101948T}).

Theoretical indication of occurrences of FFC in CCSN and BNSM implies that its feedback onto radiation-hydrodynamics and nucleosyntheisis needs to be incorporated in one way or another. Very recently, there have been some attempts to tackle this issue in simulations of BNSM remnants (see e.g., \cite{2021PhRvL.126y1101L,2022arXiv220316559J}), which certainly marked an important stepping stone towards BNSM models with quantum kinetics neutrinos. However, there are many simplifications and approximations in these models, which would discard important features of neutrino-flavor conversion. It is, hence, necessary to consider how we bridge the current gap between CCSN/BNSM simulations and non-linear dynamics of neutrino quantum kinetics. The numerical code which we present in this paper is designed to mediate their binding to each other.

In the last decades, considerable progress have also been achieved in neutrino-oscillation community. Analytic approaches with simplifying assumptions and toy models have facilitated our understanding of neutrino-flavor conversion (see, e.g., \cite{2006PhRvD..74j5010H,2007PhRvD..76l5008R,2007PhRvD..75l5005D,2007PhRvD..75h3002R,2015IJMPE..2441008D,2020PhRvD.101d3009J,2022PhRvL.128l1102P}). Numerical simulations are also powerful tools to explore their non-linear behaviors with relaxing assumptions. However, they are not yet at a stage to provide reliable astrophysical consequences of the flavor conversion. This is mainly due to the fact that there are large disparities in spatial- and temporal scales between neutrino-flavor conversion and astrophysical phenomena, exhibiting the need for currently unfeasible computational power. Notwithstanding, there are many numerical simulations to study the non-linear properties of neutrino quantum kinetics. The simplest model would be neutrino bulb model\footnote{We note that there are different levels of approximations in the light bulb model; for instances, steady-state or time-dependent, single or multi-angle, with or without including halo effects. See references for more details.} \cite{2006PhRvD..74l3004D,2006PhRvD..74j5014D,2007PhRvD..76h5013D,2007JCAP...12..010F,2011PhRvL.106i1101D,2012PhRvD..85f5008D,2017PhRvD..96b3009Y,2018PhRvD..97h3011V,2018PhRvD..98j3020Z,2021PhRvD.103f3008Z}. Although this model has many simplifying assumptions, some intriguing features of neutrino-flavor conversion in global scale have been revealed. Two-beam-  \cite{2015PhRvD..92l5030D,2019PhRvL.122i1101C} and line-beamed models \cite{2015PhRvD..92f5019A,2018PhRvD..98d3014A,2019PhLB..790..545A}, both of which reduce the computational cost by limiting neutrino flight-directions (see also \cite{2015PhRvD..92b1702M}), are also powerful approaches to study non-linear phase of flavor conversion without much computational burden. More direct simulations of flavor conversion has also been made under homogeneous- \cite{2020PhRvD.101d3009J,2020PhRvD.102j3017J,2021PhLB..82036550X,2021PhRvD.104b3011S,2021arXiv210914011S,2022PhRvL.128l1102P,2022arXiv220411873H} and inhomogeneous- \cite{2017JCAP...02..019D,2019PhRvD.100b3016M,2020PhLB..80035088M,2020PhRvD.102f3018B,2021PhRvL.126f1302B,2021PhRvD.104j3003W,2021PhRvD.103f3001M,2021PhRvD.104j3023R,2021PhRvD.104h3035Z,2021PhRvD.104l3026D,2022JCAP...03..051A,2022PhRvD.105d3005S,2022arXiv220505129B,2022arXiv220600676S} neutrino medium with resolving neutrino angular distributions in momentum space. It is also worthy of note that a code-comparison across different numerical solvers was made very recently \cite{2022arXiv220506282R}, which is a rewarding effort to convince them and others that their quantum kinetics codes work well, and to understand strengths and weaknesses of each code.

On the other hand, time-dependent features of neutrino-flavor conversions in global scale remain an enduring mystery. It should be mentioned that collective neutrino oscillations naturally break their own temporal stationarity \cite{2015PhLB..751...43A,2015PhRvD..92l5030D}, exhibiting the importance of time-dependent simulations. General relativistic (GR) effects also need to be incorporated in global simulations, since the gravity is usually strong in regions where neutrino number density is high (e.g., in the vicinity of neutron star). They may play negligible roles in flavor conversion, since gravitational redshift and light-bending effects have an influence on neutrino distributions in momentum space, that has an impact on self-interaction potentials (see, e.g. \cite{2017PhRvD..96b3009Y}). However, currently available numerical codes (see, e.g., \cite{2021PhRvD.103h3013R,2022arXiv220312866G}), that have a capability of solving time-dependent quantum kinetic equation, were designed for local simulations. More precisely speaking, their numerical approach is not suited for curvilinear coordinate system, and the formulation is not applicable to neutrino transport in curved spacetimes. As is well established in numerical treatments of GR Boltzmann equation, the transport operator can be, in general, written in a conservative form (see, e.g., \cite{2013PhRvD..88b3011C,2014ApJS..214...16N,2020ApJ...888...94D}), which is very useful for numerical simulations. The operator contains not only spatial components but also momentum-space ones (see Sec.~\ref{sec:basiceq} for more details), that accounts for geometrical effects of neutrino transport. As such, the formalism is suited for neutrino transport in global scale. Our CCSN neutrino-radiation-hydrodynamic code with full Boltzmann neutrino transport was developed with the formalism \cite{2014ApJS..214...16N,2017ApJS..229...42N,2019ApJ...878..160N,2021ApJ...909..210A}, and it has worked well in multi-dimensional (multi-D) CCSN simulations \cite{2018ApJ...854..136N,2019ApJ...880L..28N,2019ApJ...872..181H,2020ApJ...903...82I}.

In this paper, we present a new numerical code GRQKNT (General-Relativistic Quantum-Kinetics Neutrino Transport), that is designed for time-dependent local- and global simulations of neutrino-flavor conversion in CCSN and BNSM environments. At the moment, we are particularly interested in the dynamics of FFC, that seems to be the most relevant to and the largest uncertainty in the theory of CCSN and BNSM. One may wonder if those simulations are intractable by the current numerical resources. However, we can relax the computational burden by reducing the neutrino number density or neutrino Hamiltonian potentials. Since the physical scale of FFC is determined only by the self-interaction potential, the reduction of neutrino number density makes the global simulations computationally feasible. We note that results obtained from this approach should be carefully checked; for instance, the resolution study is indispensable not only for real space but also momentum one. In fact, mode-couplings in FFC generate small-scale angular structures from large angular scales (similar as turbulent cascades, see, e.g., \cite{2020PhRvD.102j3017J}), and slow neutrino-flavor conversions may induce sharp spectral-swapping in energy direction. These facts suggest that high numerical resolutions in the energy direction may be still necessary even reduction of the Hamiltonian potential. The resolution study would help us to exclude spurious evolution of flavor conversion.

It is worthy of note that the similar approach can be seen in other fields; for instance, ion-to-electron mass ratio is frequently reduced in particle-in-cell simulations of plasma physics to save computational time\footnote{It is worth to note that nowadays the increased computational resources allow PIC simulations with real mass ratio (see, e.g., \cite{2015PPCF...57k3001A}).}. Realistic FFC features (i.e., without reduction of neutrino number density) can be obtained by increasing the neutrino number density, and the resolutions in neutrino phase space and the size of computational domain are controlled in accordance with computational power. Following the above approach, we carried out a time-dependent global simulations of FFC; the results are reported in a separate paper \cite{2022arXiv220604097N}. We confine the scope of this paper to describing philosophy, design, and numerical aspects of GRQKNT.

This paper is organized as follows. We describe the basic equation and the numerical formalism in Sec.~\ref{sec:basiceq}. We encapsulate the detail of each numerical module into each section: transport module (in Sec.~\ref{sec:transport}), collision term (in Sec.~\ref{sec:Colterm}), and oscillation module (in Sec.~\ref{sec:osc}). Finally, we summarize and conclude in Sec.~\ref{sec:summary}. We use the unit with $c = G = \hbar = 1$, where $c$, $G$, and $\hbar$ are the light speed, the gravitational constant, and the reduced Planck constant, respectively. We use the metric signature of $- + + +$.

\section{Basic equations}\label{sec:basiceq}

In GRQKNT code, we solve general relativistic mean-field quantum kinetic equation (QKE), which is written as (see also \cite{2019PhRvD..99l3014R}),
\begin{equation}
\begin{aligned}
p^{\mu} \frac{\partial \barparena{f}}{\partial x^{\mu}}
+ \frac{dp^{i}}{d\tau} \frac{\partial \barparena{f}}{\partial p^{i}}
= - p^{\mu} u_{\mu} \barparena{S}
+ i p^{\mu} n_{\mu} [\barparena{H},\barparena{f}].
\end{aligned}
\label{eq:basicneutrinosQKE}
\end{equation}
In the expression, we use the same convention as \cite{2021ApJS..257...55K}\footnote{This is also the same convention that used in \cite{2019PhRvD..99l3014R}, although there is a typo in right hand side of Eq.~9 in the paper (computing self-interaction potentials). $\bar{f}^{\prime}$ needs to be replaced to $\bar{f}^{*\prime}$, which is confirmed with one of the authors (Sherwood Richers, private communication). We also note that our convention for $\bar{f}$ corresponds to $\bar{\rho}^{*}$ in \cite{2006PhRvD..74j5010H} (see, e.g., Eq.~A2 in the paper), which has been frequently used in the literature.}. $f$ and $\bar{f}$ denote the density matrix of neutrinos and anti-neutrinos, respectively; $x^{\mu}$ and $p^{\mu}$ are spacetime coordinates and the four-momentum of neutrinos (and anti-neutrinos); $u^{\mu}$ and $n^{\mu}$ represent the four-velocity of fluid and the unit vector normal to the spatial hypersurface of constant time, respectively; $S$ ($\bar{S}$) represents the collision terms measured at the fluid rest frame; $H$ ($\bar{H}$) denotes the Hamiltonian operator associated with neutrino-flavor conversion. The Hamiltonian is composed of three compositions,
\begin{equation}
\barparena{H} = \barparena{H}_{\rm vac} + \barparena{H}_{\rm mat} + \barparena{H}_{\nu \nu}, \label{eq:Hdecompose}
\end{equation}
where
\begin{equation}
\begin{aligned}
&\bar{H}_{\rm vac} = H^{*}_{\rm vac} , \\
&\bar{H}_{\rm mat} = - H^{*}_{\rm mat} ,\\
&\bar{H}_{\nu \nu} = - H^{*}_{\nu \nu}.
\end{aligned}
\label{eq:Hantineutrinos}
\end{equation}

$H_{\rm vac}$ denotes the vacuum Hamiltonian with the expression in the neutrino-flavor eigenstate, which can be written as
\begin{equation}
\begin{aligned}
H_{\rm vac} = \frac{1}{2 \nu} U 
\begin{bmatrix}
        m^{2}_{1} & 0 & 0\\
        0 & m^{2}_{2} & 0 \\
        0 & 0 & m^{2}_{3}
    \end{bmatrix}
 U^{\dagger} ,\\
\end{aligned}
\label{eq:Hvdef}
\end{equation}
where $\nu = -p^{\mu} n_{\mu} = p^{0} \alpha$; $\alpha$ denotes the lapse function associated with space-time foliation (3+1 formalism of curved space-time); $m_{i}$ denotes the mass of neutrinos; $U$ denotes the Pontecorvo-Maki-Nakagawa-Sakata (PMNS) matrix. The matter potential $H_{\rm mat}$ can be written as
\begin{equation}
\begin{aligned}
H_{\rm mat} = D
\begin{bmatrix}
        V_e & 0 & 0\\
        0 & V_{\mu} & 0 \\
        0 & 0 & V_{\tau} + V_{\mu \tau}
    \end{bmatrix}
 ,\\
\end{aligned}
\label{eq:Hmatdef}
\end{equation}
where $D = (-p^{\mu} u_{\mu})/\nu$ denotes the effective Doppler factor between the laboratory frame and the fluid-rest frame, i.e., representing the Lorentz boost between $\mbox{\boldmath $n$}$ and $\mbox{\boldmath $u$}$ under local flatness (see \cite{2014ApJS..214...16N,2017ApJS..229...42N} for more details). The leading order of $V_{\ell}$ can be written as
\begin{equation}
V_{\ell} = \sqrt{2} G_F (n_{\ell^{-}} - n_{\ell^{+}}) , \label{eq:Velldef}
\end{equation}
where
$G_F$ and $n_{\ell}$ represent the Fermi constant and the number density of charged-leptons $(\ell = e, \mu, \tau)$, respectively. As a default set, we assume that on-shell heavy leptons ($\mu$ and $\tau$) do not appear, i.e., $V_{\mu}$ and $V_{\tau}$ are set to be zero. It should be mentioned, however, that $V_{\mu}$ may not always be zero, since on-shell muons would appear in the vicinity of (or inside) neutrino star \cite[see, e.g.,][]{2017PhRvL.119x2702B,2020PhRvD.102l3001F}. $V_{\mu \tau}$ represents, on the other hand, the radiative correction of neutral current \cite{1987PhRvD..35..896B,2000PhRvD..62c3007D}, which is a leading order to distinguish $\nu_{\mu}$ and $\nu_{\tau}$ in cases with $V_{\mu}=V_{\tau}=0$. Following \cite{2000PhRvD..62c3007D}, $V_{\mu \tau}$ can be computed as,
\begin{equation}
V_{\mu \tau} = V_e \frac{3 G_F m_{\tau}^2 }{2 \sqrt{2} \pi^2 Y_e} \left( \ln \frac{m_{W}^2}{m_{\tau}^{2}} - 1 + \frac{Y_n}{3}  \right), \label{eq:Vmyutau}
\end{equation}
where $m_{\tau}$ and $m_{W}$ denote the mass of tau and W boson, respectively. $Y_e$ and $Y_n$ represents the electron-fraction and neutron-fraction, respectively.

$H_{\nu \nu}$ represents the neutrino self-interaction potential, which can be written as
\begin{equation}
H_{\nu \nu} = \sqrt{2} G_F \int \frac{d^3 q^{\prime}}{(2 \pi)^3} (1 - \sum_{i=1}^{3} \ell^{\prime}_{(i)} \ell_{(i)} ) (f(q^{\prime}) - \bar{f}^{*}(q^{\prime})), \label{eq:Hselfpotedef}
\end{equation}
where $d^3q$ denotes the momentum space volume of neutrinos, which are measured at the laboratory frame; $\ell_{i} (i = 1, 2, 3)$ denote directional cosines for the direction of neutrino propagation. The two angles of neutrino flight directions are measured with respect to a spatial tetrad basis $\mbox{\boldmath $e$}_{(1)}$. There are multiple options to choose $\mbox{\boldmath $e$}_{(1)}$, and we usually set it as a unit vector in the same direction of the radial coordinate basis (see e.g., \cite{2014PhRvD..89h4073S,2017ApJS..229...42N}). By using the polar- ($\theta_{\nu}$) and azimuthal angles ($\phi_{\nu}$) in neutrino momentum space, $\ell_{i} (i = 1, 2, 3)$ can be written as
\begin{equation}
  \begin{split}
&\ell_{(1)} = \cos \hspace{0.5mm} \theta_{\nu}, \\
&\ell_{(2)} = \sin \hspace{0.5mm} \theta_{\nu}   \cos \hspace{0.5mm} \phi_{\nu}, \\
&\ell_{(3)} = \sin \hspace{0.5mm} \theta_{\nu}   \sin \hspace{0.5mm} \phi_{\nu}.
  \end{split}
\label{eq:el}
\end{equation}

There are four remarks regarding the QKE. First, we take the relativistic limit of neutrinos in the expression; the energy of neutrinos is much larger than the rest-mass energy, which is a reasonable approximation for CCSN and BNSM\footnote{The typical energy of neutrinos in CCSN and BNSM is an order of $10$ MeV, meanwhile the current upper bound of neutrino mass is $\lesssim 0.1$ eV \cite{2019PhRvL.123h1301L}.}. Hence, we treat the neutrinos as massless particles in the transport equation (the left hand side of Eq.~\ref{eq:basicneutrinosQKE}) and the collision term (the first term in the right hand side of Eq.~\ref{eq:basicneutrinosQKE}). On the other hand, we leave the leading term of $\nu \times (m/\nu)^2$ in the Hamiltonian operator (see Eq.~\ref{eq:Hvdef}). Second, we define the Hamiltonian operator in the laboratory frame, although the choice of the frame is arbitrary (see, e.g., \cite{2019PhRvD..99l3014R}). Third, GRQKNT code is also compatible with two-flavor approximations. In simulations under the two-flavor approximation, we can change the size of density matrix and Hamiltonian operators from $3 \times 3$ to $2 \times 2$ in GRQKNT. In the two-flavor case, the vacuum oscillation parameters are also changed, which is determined according to the problem. Fourth, Eq.~\ref{eq:basicneutrinosQKE} corresponds to the mean-field approximation or one-body density matrix description with the first truncation of BBGKY hierarchy (see \cite{2015IJMPE..2441009V} for more details). Under the assumption, the neutrino self-interaction is treated as an interaction between each neutrino and their mean-field neutrino medium in its vicinity. However, there may be astrophysical regimes where mean-field approximation is inappropriate (see e.g., \cite{2019PhRvD..99l3013P,2019PhRvD.100h3001C}) and may lead to different astrophysical consequence \cite{2018PhRvD..98h3002B}. Thus, GRQKNT is not capable of capturing all features of neutrino-flavor conversion. We leave the task incorporating these many-body corrections into GRQKNT to future work.

Following \cite{2014PhRvD..89h4073S}, we cast the QKE in a conservative form. This is a useful formalism for numerical simulations, since the neutrino-number conservation can be ensured up to machine-precision. This can be written as,
\begin{equation}
  \begin{split}
&\frac{1}{\sqrt{-g}} \left. \frac{\partial}{\partial x^{\alpha}} \right|_{q_{i}}
\Biggl[  \Bigl( n^{\alpha} + \sum^{3}_{i=1} \ell_{i} e^{\alpha}_{(i)} \Bigr) \sqrt{-g} \barparena{f}   \Biggr] \\
& - \frac{1}{\nu^2} \frac{\partial}{\partial \nu}( \nu^3 \barparena{f} \omega_{(0)}  )
+ \frac{1}{\sin\theta_{\nu}} \frac{\partial}{\partial \theta_{\nu}}
( \sin\theta_{\nu} \barparena{f} \omega_{(\theta_{\nu})} ) \\
& + \frac{1}{ \sin^2 \theta_{\nu}} \frac{\partial}{\partial \phi_{\nu}} (\barparena{f} \omega_{(\phi_{\nu})}) = D \barparena{S} - i [\barparena{H},\barparena{f}],
  \end{split}
\label{eq:conformQKE}
\end{equation}
where $g, x^{\alpha}$ are the determinant of the four-dimensional metric, coordinates of spacetime, respectively. $e^{\alpha}_{(i)} (i = 1, 2, 3)$ denote a set of the (spatial) tetrad bases normal to $n$. The factors of $\omega_{(0)}, \omega_{(\theta_{\nu})}, \omega_{(\phi_{\nu})}$ are given as,
\begin{equation}
  \begin{split}
& \omega_{(0)} \equiv \nu^{-2} p^{\alpha} p_{\beta} \nabla_{\alpha} n^{\beta}, \\
& \omega_{(\theta_{\nu})} \equiv \sum^{3}_{i=1} \omega_{i} \frac{ \partial \ell_{(i)} }{\partial \theta_{\nu} }, \\
& \omega_{(\phi_{\nu})} \equiv \sum^{3}_{i=2} \omega_{i} \frac{ \partial \ell_{(i)} }{\partial \phi_{\nu} }, \\
&\omega_{i} \equiv \nu^{-2} p^{\alpha} p_{\beta} \nabla_{\alpha} e^{\beta}_{(i)},
  \end{split}
\label{eq:Omega}
\end{equation}
which can also be expressed with the Ricci rotation coefficients \cite{2014PhRvD..89h4073S}. Spherical polar coordinate is often employed in solving Boltzmann equation, and we chose a set of tetrad basis, $\mbox{\boldmath $e$}_{(i)}$, having the following coordinate components,
\begin{equation}
  \begin{split}
& e^{\alpha}_{(1)} = (0, \gamma^{-1/2}_{rr}, 0, 0 ) \\
& e^{\alpha}_{(2)} = \Biggl(0, -\frac{\gamma^{-1/2}_{r \theta}}{\sqrt{\gamma_{rr} (\gamma_{rr} \gamma_{\theta \theta} - \gamma^2_{r \theta})}}, \sqrt{ \frac{\gamma_{rr}}{ \gamma_{rr} \gamma_{\theta \theta} - \gamma^2_{r \theta} } }, 0 \Biggr) \\
& e^{\alpha}_{(3)} = \Biggl(0, \frac{\gamma^{r \phi}}{\sqrt{\gamma^{\phi \phi}}} , \frac{\gamma^{\theta \phi}}{\sqrt{\gamma^{\phi \phi}}}, \sqrt{\gamma^{\phi \phi}} \Biggr),
  \end{split}
\label{eq:polartetrad}
\end{equation}
where $\mbox{\boldmath $\gamma$}$ denotes the induced metric on each spatial hypersurface.

Here, we explicitly write down the QKE in flat spacetime with spherical polar coordinate, which is also useful to see geometrical effects. Eq.~\ref{eq:conformQKE} can be rewritten in flat spacetime as,
\begin{equation}
  \begin{split}
& \frac{\partial \barparena{f}}{\partial t}
+ \frac{1}{r^2} \frac{\partial}{\partial r} ( r^2 \cos \theta_{\nu}  \barparena{f} )
+ \frac{1}{r \sin \theta} \frac{\partial}{\partial \theta} (  \sin \theta \sin \theta_{\nu} \cos \phi_{\nu} \barparena{f}   ) \\
& + \frac{1}{r \sin \theta} \frac{\partial}{\partial \phi} ( \sin \theta_{\nu} \sin \phi_{\nu} \barparena{f} ) - \frac{1}{r \sin \theta_{\nu}} \frac{\partial}{\partial \theta_{\nu}} ( \sin^2 \theta_{\nu} \barparena{f}) \\
& - \frac{\cot \theta}{r}  \frac{\partial}{\partial \phi_{\nu}} ( \sin \theta_{\nu} \sin \phi_{\nu} \barparena{f} )  = D \barparena{S} - i [\barparena{H},\barparena{f}].
  \end{split}
\label{eq:flatQKE}
\end{equation}
In order to see differences from QKE with Cartesian coordinate, there are two points to which attention should be paid. First, Jacobian determinant of three-dimensional real space ($r^2 \sin \theta$) appears in the spatial transport terms (the second to fourth terms in the left hand side of Eq.~\ref{eq:flatQKE}), which is directly related to $\sqrt{-g}$ in Eq.~\ref{eq:conformQKE}. Second, Eq.~\ref{eq:flatQKE} has transport terms in momentum space (the fifth- and sixth terms in the left hand side of the equation). This is attributed to the fact that $\mbox{\boldmath $e$}_{(i)}$ is not spatially uniform but rather rotates with $\theta$. As a result, $\omega_{i}$ becomes non-zero values even in the flat-spacetime (see in Eq.~\ref{eq:Omega}). The neutrino advection in angular directions of momentum space can also be interpreted more intuitively as follows. Neutrinos traveling straight in space experience different directional cosine with respect to $\mbox{\boldmath $e$}_{(1)}$ except for those propagating in the same direction with $\mbox{\boldmath $e$}_{(1)}$. The out-going neutrinos with a finite angle with $\mbox{\boldmath $e$}_{(1)}$ becomes more forward peaked with increasing radius. This is essentially the same as geometrical effects that have also been discussed in light bulb model. In fact, light bulb model can be restored by solving Eq.~\ref{eq:flatQKE} in spherically symmetry with injected outgoing neutrinos from a certain radius.

It is worthy of note that Eq.~\ref{eq:flatQKE} does not compromise the applicability of GRQKNT to local simulations. As mentioned above, transport terms in angular directions of momentum space are associated with variations of coordinate basis, indicating that they are negligible if we make the simulation box enough small so that the coordinate curvature can be safely neglected. In this case, Jacobian determinant appearing in spatial transport terms can also be dropped, exhibiting that QKE with Cartesian coordinate is restored. Here, we provide an example of the numerical setup. Let us consider a three-dimensional box in space with a region of $R \le r \le R+\Delta R$, $\Theta - \Delta \theta/2 \le \theta \le \Theta + \Delta \theta/2$, and $\Phi - \Delta \phi/2 \le \phi \le \Phi + \Delta \phi/2$ in radial, zenith- and azimuthal direction, respectively. The simulation box becomes a cubic shape when we choose a set of parameters as $\Delta R/R = \Delta \theta = \Delta \phi \ll 1$, $\Theta = \pi/2$, and $\Phi = 0$. In Sec.~\ref{sec:FFC}, we demonstrate 1D local simulations by following this numerical setup.

In the current version of GRQKNT code, we can run simulations of neutrino-flavor conversion in three representative spacetimes: flat spacetime, Schwarzschild black hole, and Kerr black hole. In Schwarzschild spacetime, we employ the Schwarzschild coordinate. The line element can be written as,
\begin{equation}
  \begin{split}
ds^2 =& - \Bigl(1 - \frac{2M}{r} \Bigr) dt^2 + \Bigl(1 - \frac{2M}{r} \Bigr)^{-1} dr^2 \\
& + r^2 d\theta^2 + r^2 \sin^2 \theta d\phi^2,
  \end{split}
\label{eq:lineSchw}
\end{equation}
where $M$ denotes the black hole mass. By using a set of tetrad basis described in Eq.~\ref{eq:polartetrad}, the resultant QKE can be written as,
\begin{equation}
  \begin{split}
& \frac{\partial }{\partial t} \biggl[ \Bigl(1 - \frac{2M}{r} \Bigr)^{-1/2} \barparena{f} \biggr]
+ \frac{1}{r^2} \frac{\partial}{\partial r} \biggl[ r^2 \cos \theta_{\nu} \Bigl(1 - \frac{2M}{r} \Bigr)^{1/2}   \barparena{f} \biggr] \\
& + \frac{1}{r \sin \theta} \frac{\partial}{\partial \theta} (  \sin \theta \sin \theta_{\nu} \cos \phi_{\nu} \barparena{f}   ) \\
& + \frac{1}{r \sin \theta} \frac{\partial}{\partial \phi} ( \sin \theta_{\nu} \sin \phi_{\nu} \barparena{f} ) \\
& - \frac{1}{\nu^2} \frac{\partial}{\partial \nu} \biggl[ \frac{M}{r^2} \Bigl( 1 - \frac{2M}{r} \Bigr)^{-1/2} \nu^3 \cos \theta_{\nu} \barparena{f} \biggr] \\
&- \frac{1}{\sin \theta_{\nu}} \frac{\partial}{\partial \theta_{\nu}} \biggl[ \sin^2 \theta_{\nu} \frac{r-3M}{r^2} \Bigl(1 - \frac{2M}{r} \Bigr)^{-1/2}   \barparena{f} \biggl] \\
& - \frac{\cot \theta}{r}  \frac{\partial}{\partial \phi_{\nu}} ( \sin \theta_{\nu} \sin \phi_{\nu} \barparena{f} )  = D \barparena{S} - i [\barparena{H},\barparena{f}].
  \end{split}
\label{eq:SchQKE}
\end{equation}
We note that Eq.~\ref{eq:SchQKE} becomes identical to that in flat-spacetimes (Eq.~\ref{eq:flatQKE}) by taking the limit of $M \to 0$.

In Kerr spacetimes, we employ Kerr-Schild coordinates; the line element can be written as,
\begin{equation}
  \begin{split}
ds^2 =& - \Bigl(1 - \frac{2Mr}{\Sigma} \Bigr) dt^2 + \frac{4Mr}{\Sigma} dt dr + \Bigl(1 + \frac{2Mr}{\Sigma} \Bigr) dr^2 \\
& + \Sigma d\theta^2 + \frac{\sin^2 \theta}{\Sigma} \biggl[  (r^2 + a^2)^2 - \Delta a^2 \sin^2 \theta   \biggr] d\phi^2 \\
& - 2 a \sin^2 \theta \Bigl(1 + \frac{2Mr}{\Sigma} \Bigr) d\phi dr \\
& - \frac{4Mar}{\Sigma} \sin^2 \theta d\phi dt,
  \end{split}
\label{eq:lineKerrSchild}
\end{equation}
where
\begin{equation}
  \begin{split}
& \Delta \equiv r^2 - 2Mr + a^2, \\
& \Sigma \equiv r^2 + a^2 \cos^2 \theta, \\
  \end{split}
\label{eq:kerrvali}
\end{equation}
and $a$ denotes kerr parameter (angular momentum of black hole per unit mass). The explicit description of conservative form of QKE is quite lengthy; hence we omit to explicitly write down here. However, it can be straightforwardly derived from Eqs.~\ref{eq:conformQKE}-\ref{eq:polartetrad} by following the procedure outlined in this section.

\section{Transport module}\label{sec:transport}


\subsection{Design}\label{sec:des_transport}
Consistent treatments of transport- and collision terms in multi-D Boltzmann neutrino transport was a technical challenge for discrete-ordinate Sn method. As described in Sec. 2 and 3 of \cite{2014ApJS..214...16N}, the main source of difficulty is solving neutrino transport while interacting with moving-matter through iso-energetic scatterings. The problem was, however, resolved by a mixed-frame approach with a two-energy-grid technique: Lagrangian-remapping grid (LRG) and Laboratory-fixed grid (LFG).

The LRG is constructed so that it does not depend on neutrino angles measured in the fluid rest frame. Thanks to the isotropy, distribution functions defined on the LRG can be used to evaluate collision terms, in particular for iso-energetic scatterings, without any problems. On the other hand, advection terms of Boltzmann equation are handled in the laboratory frame, implying that we need Lorentz transformations from the fluid rest frame to the laboratory one. Consequently, LRG becomes angular-dependent in the Laboratory frame. It should also be mentioned that the LRG depends on space in general (since the fluid velocity is not uniform), which is inconvenient to evaluate transport operators. We, hence, define LFG, which is constructed so that it does not depend on both neutrino angles, space, and time in Laboratory frame. After interpolating distribution functions from LRG to LFG, we evaluate the neutrino flux at (spatial) cell-boundaries on LFG. We then calculate the numerical flux on each LRG by taking into account the overlapped region with respect to LFGs. As a result, we can consistently handle both advection and collision terms. This technique has already been extended in cases with curved spacetimes \cite{2017ApJS..229...42N,2021ApJ...909..210A}; hence, we adopt the same technique in GRQKNT code.

As described in Sec.~\ref{sec:basiceq}, we choose $\mbox{\boldmath $n$}$ as a tetrad basis in transport terms of QKE, exhibiting that the neutrino transport is solved in the laboratory frame. We use LFG to evaluate the neutrino advection in both space and momentum space except for energy direction\footnote{We handle the neutrino advection in energy direction by LRG, since no technical issues arise in LRG.}. It should be mentioned that we employ an explicit time-integration scheme in GRQKNT, whereas a semi-implicit method is implemented in our Boltzmann code \cite{2014ApJS..214...16N}. This suggests that the implementation of neutrino advection in GRQKNT code is much simpler than the Boltzmann solver. More precisely speaking, computations of matrix inversion are not involved in GRQKNT. We also note that the two-energy-grid technique is not necessary in cases with neglecting fluid-velocities or gray neutrino transport (i.e., energy-integrated QKE). In these cases, the LFG is set to be the same as LRG. We refer readers \cite{2014ApJS..214...16N} for the detail of numerical implementation of the two-energy-grid technique.

Except for the two-energy-grid technique, solving Eq.~\ref{eq:conformQKE} is quite straightforward. We apply a well-established hyperbolic solver, 5th-order weighted essentially non-oscillatory or WENO scheme \cite{1994JCoPh.115..200L} with some extensions. The WENO scheme is easy to implement and run multi-D simulations; in fact, another neutrino-flavor conversion solver recently developed in \cite{2022arXiv220312866G} also employs 7th-order WENO. Below, we describe the WENO scheme implemented in our GRQKNT code.

The numerical implementation of WENO is essentially the same way used in \cite{2018AcMSn..34...37H}. The advantage of this method is a simple and computationally-cheap implementation of WENO on non-uniform grids. It should be mentioned that non-uniform grids are frequently used in global simulations of CCSN and BNSM, since they are useful to reduce computational costs without compromising accuracy. On the other hand, WENO schemes on non-uniform grids requires, in general, computations of complicated weight-functions (see, e.g., \cite{2008JCoPh.227.2977C}), causing the increase of CPU-time. In \cite{2018AcMSn..34...37H}, they proposed a new method to save the CPU time with sustaining the accuracy; we hence adopt the method in our GRQKNT code\footnote{There is a caveat, however. Highly non-uniform grids would reduce the accuracy of solver to second-order, which is a weak point in the method of \cite{2018AcMSn..34...37H}. However, such highly non-uniform grids are not necessary in simulations in CCSN and BNSM environments; hence this limitation does not compromise the usability of GRQKNT code.}.

For the basic part of the WENO scheme, we refer readers to \cite{2018AcMSn..34...37H}, and we only describe two extensions from the original one. First, we implement a five-stage fourth-order strong stability-preserving TVD Runge-Kutta by following \cite{spiteri2002new,2020JOUC...19..747G}. It should be mentioned that a fourth-order is a requirement to follow the time evolution of collective neutrino oscillations (see \cite{2021PhRvD.103h3013R,2021ApJS..257...55K}). Second, we change to compute weight function $\Omega_k$ that is used to reconstruct a physical quantity at each cell interface\footnote{In our GRQKNT code, the primitive variable corresponds to each matrix element of $\barparena{f}$.}. The $\Omega_{k}$ is defined as,
\begin{equation}
\Omega_k = \frac{\alpha_k}{\alpha_0 + \alpha_1 + \alpha_2}.
\label{eq:compomega_k}
\end{equation}
where $k$ runs from 0 to 1. In the original WENO scheme, $\alpha_k$ is computed as,
\begin{equation}
\alpha_k = \frac{C_k}{(\epsilon + IS_k)^p},
\label{eq:compalpha_k_orig}
\end{equation}
with $\epsilon=10^{-6}$ and $p=2$. $C_k$ are $C_0 = 0.1, C_1=0.6,$ and $C_2=0.3$. $IS_k$ denotes a smoothness measure. The explicit description of $IS_k$ can be seen in Eqs.~10, 12, and 15 in of \cite{2018AcMSn..34...37H}. In our test computations, however, we find that the weight function is not sufficient to sustain the stability. This is mainly due to the fact that density matrix of neutrinos has, in general, order-of-magnitude variations, which has been frequently observed CCSN simulations with full Boltzmann neutrino transport. In such a large variation, $\epsilon$ does not work well as a limiter to determine $\alpha_k$; consequently, it leads to numerical instabilities. We resolve the issue by introducing a normalization factor Q and another limiter $\epsilon_2$ to evaluate $\alpha_k$, which are
\begin{equation}
\alpha_k = \frac{C_k}{Z},
\label{eq:compalpha_k_ours}
\end{equation}
where 
\begin{equation}
  \begin{split}
&Z = \max \biggl( \epsilon_2, (\epsilon \hspace{0.5mm} Q + IS_k)^p \biggr), \\
&Q = \frac{(|q^0| + |q^1| + |q^3|)^2}{9}. 
  \end{split}
\label{eq:Qequiv}
\end{equation}
$q^k$ denotes the interfacial states of physical quantity (i.e., each element of density matrix of neutrinos); the exact expression can be seen in Sec. 2.2 of \cite{2018AcMSn..34...37H}. $Q$ corresponds to a normalization factor to make a limiter $\epsilon$ work properly. It should be noted that both Q and $IS_k$ become null when the density matrix is zero everywhere, leading to division by zero in computations of $\alpha_k$. We thus introduce another limiter $\epsilon_2$, which is set as $=10^{-50}$.

\subsection{Code test}\label{sec:tests_transport}

\begin{figure*}
   \includegraphics[width=\linewidth]{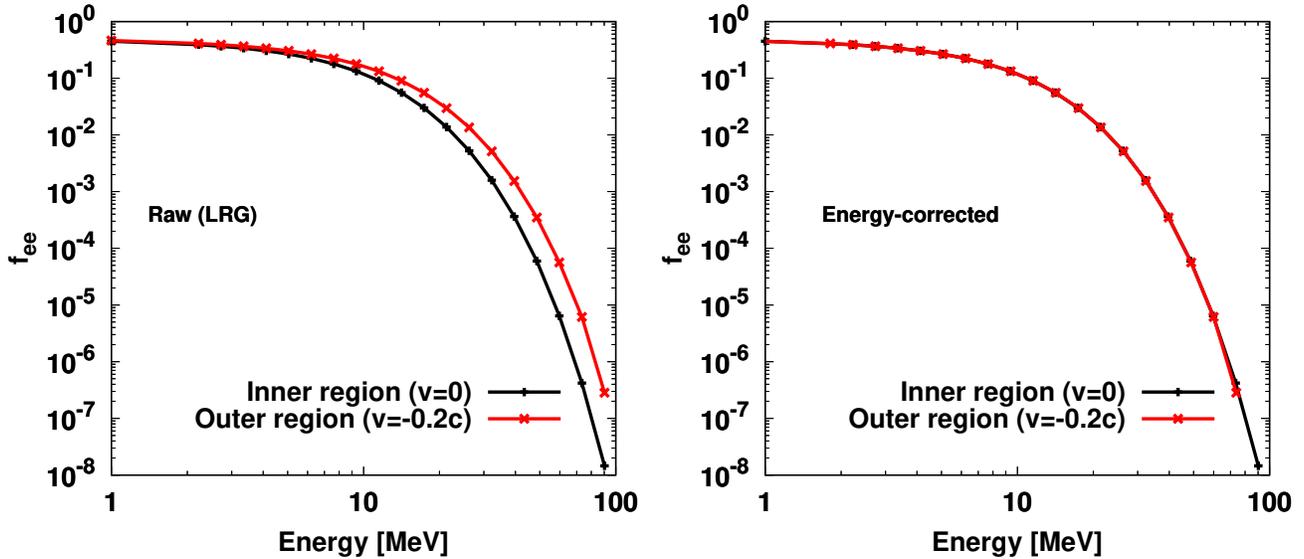}
   \caption{Transport test to check the capability of two-energy-grid technique. Left: comparison of energy-spectrum of outgoing neutrinos ($\cos \theta_{\nu}=1$) between inner- (black line) and outer region (red). The neutrinos are emitted from the matter at rest in the inner region. In the outer region, we assume that the fluid has a velocity of $- 0.2 c$, where $c$ denotes the speed of light. Since the LRG is defined on the fluid rest frame, the energy-spectrum is blue shifted. Right: Same as the left panel but we correct the neutrino energy by Doppler factor, i.e., measuring spectra in Laboratory frame. The two spectra are well-matched each other, exhibiting the fluid-velocity dependence is properly handled. See text for more details.
}
   \label{graph_LagLabo}
\end{figure*}

\begin{figure}
   \includegraphics[width=\linewidth]{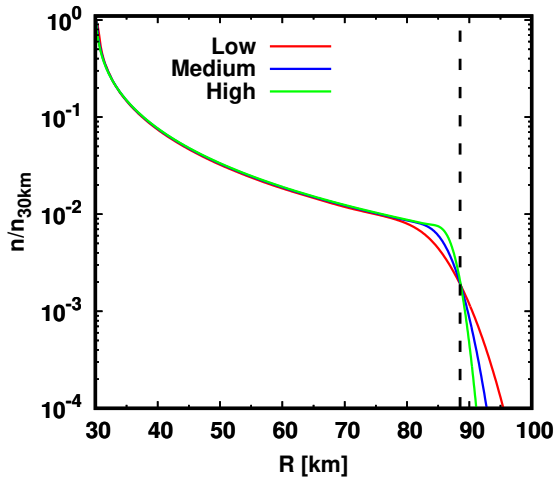}
   \caption{Transport test in Kerr-spacetime. The plots show number density of neutrinos as a function of radius at $t = 0.6$ ms. The vertical axis is normalized by that at the inner boundary ($R=30 {\rm km}$). Color distinguishes different resolutions: Low (red), Medium (blue), and High (green). The vertical black dashed-line denotes the radius where the initially-injected neutrinos reaches, which is obtained by solving a geodesic equation; see text for details.
}
   \label{graph_Radi_vs_Numdenratio_kerrcheck}
\end{figure}

\begin{figure*}
   \includegraphics[width=\linewidth]{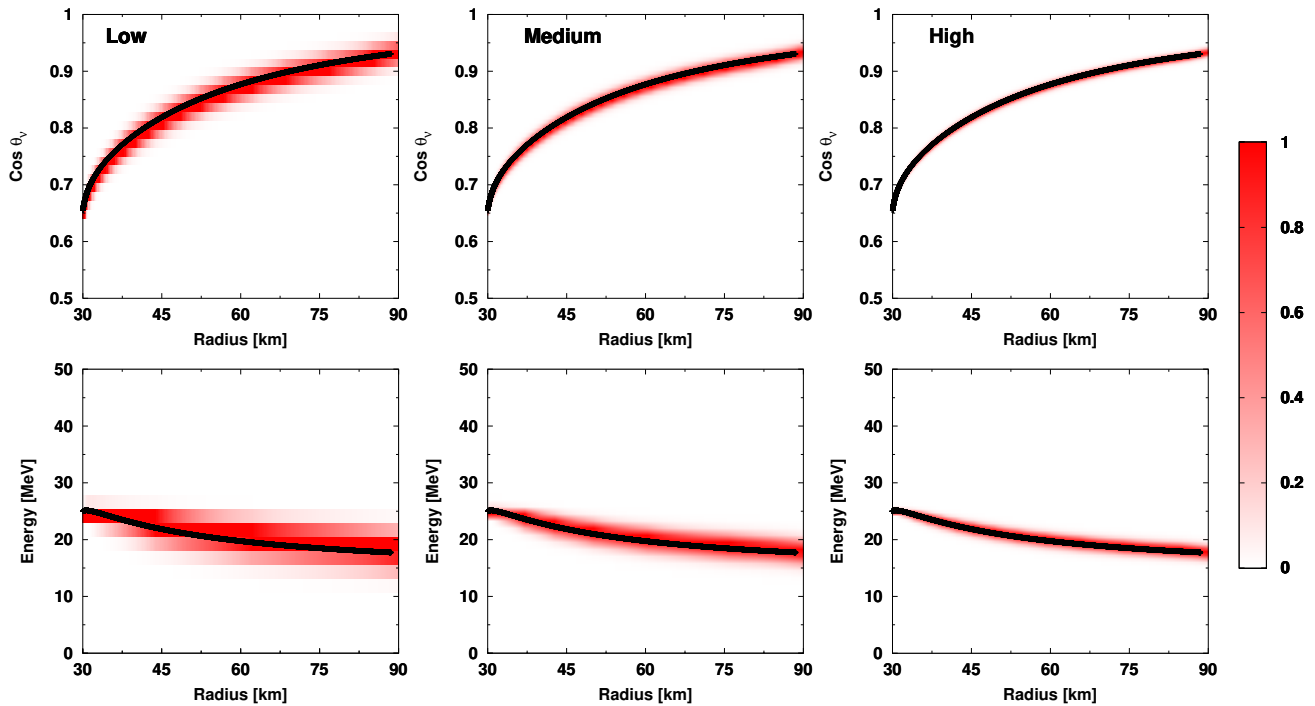}
   \caption{Transport test in Kerr-spacetime. Top: radius vs. neutrino angles (in momentum space). Bottom: radius vs. neutrino energy. Color denotes the energy-integrated $f_{ee}$ normalized by their maximum at each radius. The black solid line denotes the neutrino trajectory obtained by solving geodesic equation. From left to right, low-, medium-, and high resolutions.
}
   \label{graph_Radi_vs_angular_kerrcheck}
\end{figure*}

\begin{figure}
   \includegraphics[width=\linewidth]{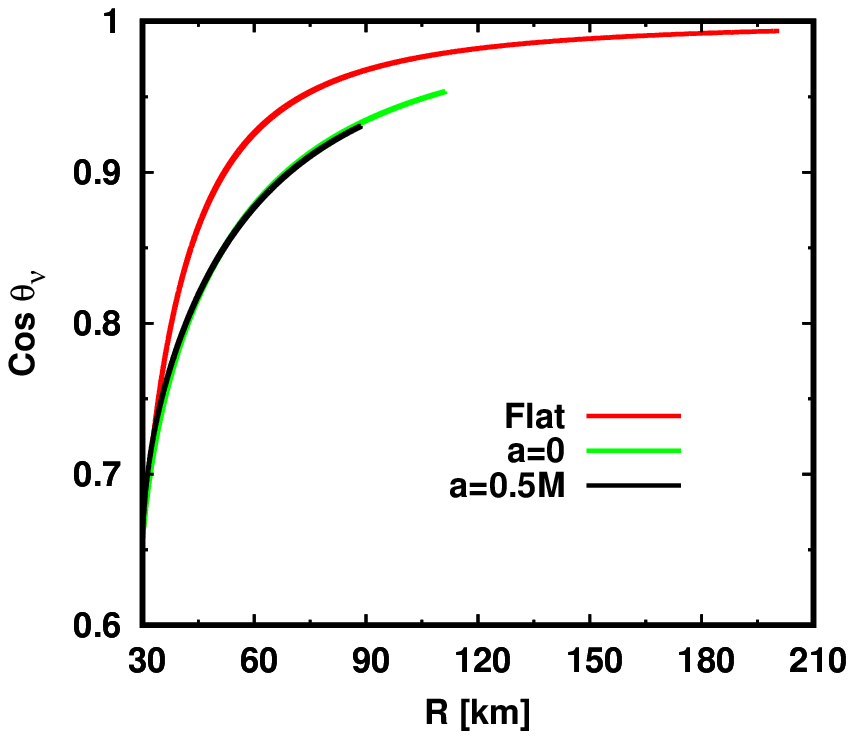}
   \caption{Radius vs. neutrino angles (in momentum space) for three different space-times: flat (red), Schwarzschild black hole (green), and Kerr black hole (black). To make a fair comparison, we use a Kerr-Schild coordinate with $a=0$ for Schwarzschild spacetime. The results are obtained by computing geodesic equations. We show each line up to the radius where the neutrino can reach by $t= 0.6$ ms.
}
   \label{graph_GeoComp}
\end{figure}

We present results of some basic tests to assess capabilities of the transport module in GRQKNT. In this test, we set collision- and oscillation terms as zero. As a result, the transport equation becomes identical among all species of neutrinos; thus, we only focus on $\nu_e$ in this section. We compare the results to those obtained by solving geodesic equation, which provides the neutrino trajectory in phase space. It should be mentioned that we check more complicated situations in this paper, in which neutrino transport and fast neutrino-flavor conversion (FFC) are coupling each other. We shall discuss the detail of the test in Sec.~\ref{sec:osc}.

We carried out a suite of transport tests in flat-spacetimes, Schwarzschild black hole, and Kerr black hole. Here, we present only the essentials with focusing on two novelties compared to other schemes: two-energy-grid technique and neutrino transport in Kerr spacetime. The tests are essentially the same as those carried out in our previous studies \cite{2014ApJS..214...16N,2021ApJ...909..210A}; hence, we refer readers them for more details.

To check the numerical implementation of two-energy grid, we inject neutrinos in the out-going direction ($\cos \theta_{\nu} = 1$) at a certain radius where the fluid is at rest. The energy spectrum of injected neutrinos is assumed to be Fermi-Dirac distribution with zero chemical potential and temperature of $5$~MeV. At the outer region, the fluid has a radial velocity of $- 0.2 c$, where $c$ denotes the speed of light. We set a discontinuity of fluid-velocity in the middle of the computational domain. It should be noted that, since LRG is defined so as to be the energy mesh becomes isotropic in the fluid-rest frame, the energy-spectrum should be shifted on LRG. We carry out a spherically symmetric simulation with $20$ energy grids, where it is discretized from 1 MeV to 100 MeV logarithmically. We employ 12 radial grid points in this simulation, i.e., the discontinuity of fluid velocity is located at the cell edge of $6$-th radial grid point. In the left panel of Fig.~\ref{graph_LagLabo}, we show the energy-spectrum on the LRG at the inner- and outer region. As expected, the energy-shift of the spectrum is confirmed\footnote{Fluid-rest-frame to Laboratory-frame transformation of energy spectrum is straightforward. The neutrino energies at laboratory- and fluid-rest frames can be transformed by one into another by Doppler factor. We also note that $f_{ee}$ is Lorentz scalar. See \cite{2014ApJS..214...16N} for more details.}. In the right panel, on the other hand, we show the energy spectrum measured at laboratory frame. This panel exhibits that the energy spectrum is good agreement with the injected energy, illustrating that the two-energy-grid technique works well.

We now turn our attention to neutrino transport in Kerr spacetime. We set $M=5 M_{\rm sun}$ and $a = 0.5 M$, where $M_{\rm sun}$ denotes the mass of the sun. Neutrinos are injected from a certain radius on equatorial plane with specifying a flight direction so as to be bounded in equatorial plane. This computational setup makes the simulation 1 (time) + 1 (radial direction in space) + 2 ($\theta_{\nu}$ and $\nu$ in momentum space) problem (see also \cite{2021ApJ...909..210A}). We chose $\phi_{\nu} = 3 \pi/2$ to maximize the frame-dragging effect of Kerr black hole. The radius, $\theta_{\nu}$-direction, and energy for injected neutrinos are assumed to be $r=30 {\rm km}$, $\cos{\theta_{\nu}}=0.655$, and $\nu=25$~MeV, respectively\footnote{It should be mentioned that the injected neutrino energy is not exactly monochromatic, and similarly there is also a finite width in the angular distribution of neutrinos. It is due to the finite-volume method in GRQKNT. Although this is one of the source of errors in the comparison to results obtained by solving geodesic equation, the deviation should be reduced with increasing resolutions in momentum space.}.

In these tests, we solve the neutrino transport in a spatial region of $30 {\rm km} \le r \le 100 {\rm km}$, where $r$ denotes the radius measured with a coordinate basis\footnote{In general, it is necessary to determine a yardstick to measure the spatial scale in curved-spacetimes. In this paper, we measure it based on a coordinate basis used in each simulation.}. We deploy uniformly $N_{r}$ grid points in the radial direction. Neutrino angular direction $\theta_{\nu}$ (lateral angular direction in momentum space) is discretized uniformly by $N_{\theta_{\nu}}$ grid points with respect to the cosine of the angle from $0^{\circ} \le \theta_{\nu} \le 180^{\circ}$. The energy-grid is also discretized uniformly by $N_{\nu}$ grid points from the range of $0 {\rm MeV} \le \nu \le 50 {\rm MeV}$. The simulations are performed at three different resolutions: low $288 (N_r) \times 128(N_{\theta_{\nu}}) \times 20 (N_{\nu})$, medium $576 \times 256 \times 40$, and high $1152 \times 512 \times 80$.

In Fig.~\ref{graph_Radi_vs_Numdenratio_kerrcheck}, we show the number density of neutrinos, measured in the laboratory frame, as a function of radius at $t=0.6$ ms. To compare the results on equal footing among three models, we normalize the density by that at $r=30$ km\footnote{In this test, we constantly inject neutrinos in time by setting $f_{ee}=0.1$ on the corresponding grid point in momentum space at the inner boundary. Since both angular- and energy grids in neutrino momentum space are not identical among different resolution models, the number density is also varied. We, hence, normalize the density by that at $r=30$ km.}. Two important conclusions can be derived from Fig.~\ref{graph_Radi_vs_Numdenratio_kerrcheck}. First, our transport module does not suffer from any numerical oscillations. It is emphasized that the stable simulation is not trivial, since the problem involves strong discontinuities both in real space and momentum space. Neutrinos are injected at a certain grid point in phase space, indicating that there are strong discontinuities of $f_{ee}$ distributions in its vicinity. Numerical viscosity plays a role for the stabilization; in fact, we find some numerical diffusions around the neutrino front-position (see $r \sim 90$ km in Fig.~\ref{graph_Radi_vs_Numdenratio_kerrcheck}). Our result suggests that a limiter of our WENO scheme works properly. In Fig.~\ref{graph_Radi_vs_Numdenratio_kerrcheck}, we also compare the result to the geodesic equation. The forefront radius of neutrinos at $t=0.6$ ms, which is $r = 88.5$ km, is displayed as a vertical dashed-line in Fig.~\ref{graph_Radi_vs_Numdenratio_kerrcheck}. We confirm that this is consistent with our results, and that the deviation decreases with increasing resolutions.

To see the resolution dependence of neutrino distributions in momentum space, we display two different color maps of $f_{ee}$ in Fig.~\ref{graph_Radi_vs_angular_kerrcheck}. In the top, we show the energy-integrated $f_{ee}$ as functions of radius and neutrino-angle ($\theta_{\nu}$). In the bottom panels, we display the angular-integrated $f_{ee}$ as functions of radius and neutrino-energy ($\nu$). For visualization purpose, we normalize $f_{ee}$ by its maximum over all angles (top panels) or energy (bottom panels) at the same radius. We find that the neutrino trajectory obtained by GRQKNT simulations is good agreement with that obtained by solving geodesic equation (black lines in each panel). We confirm that numerical diffusions occur in both angular- and energy- directions, which are reduced with increasing resolutions. These results exhibit correct implementation of neutrino transport in our code.

Before closing this section, we put some remarks on effects of curved spacetimes. Aside from gravitational redshift (as shown in bottom panels of Fig.~\ref{graph_Radi_vs_angular_kerrcheck}), there are remarkable effects of curved spacetimes on the neutrino angular advection, that can be seen in Fig.~\ref{graph_GeoComp}. By comparing to the case with flat spacetime (red line), the neutrinos angular distributions are less forward-peaked in black hole spacetimes. We also note that the forefront radius of neutrinos at $t=0.6$ ms strongly depends on the choice of spacetimes. Since we inject the neutrino in the retrograde direction with respect to angular momentum of Kerr black hole, the frame-dragging effect add an attractive force (gravity becomes effectively strong); consequently the forefront radius of neutrinos becomes smaller than the case with the same  mass of Schwarzschild black hole. As shown in Fig.~\ref{graph_Radi_vs_Numdenratio_kerrcheck}, the neutrino forefront radii obtained by our simulations are good agreement with that obtained by geodesic equation, supporting that frame-dragging effects are also correctly captured in our transport module.

\section{Collision term}\label{sec:Colterm}
Neutrino-matter interactions contributing to the collisional term ($S$ in Eq.~\ref{eq:basicneutrinosQKE}) naturally change neutrino distributions in momentum space. It has been demonstrated that the interplay between neutrino transport and collision term leads to energy-, angle-, and flavor dependent neutrino dynamics in CCSN and BNSM. There remain, however, large uncertainties in rolls of collision term in neutrino-flavor conversion. This issue attracted a great deal of attention, and some important progress has been made in the last decade. The collisional instability proposed by \cite{2021arXiv210411369J} represents a case that collision term directly drives flavor conversions. The so-called neutrino halo effect, that is associated with momentum-exchanged scatterings, potentially plays a dominant role to induce slow neutrino-flavor conversion \cite{2012PhRvL.108z1104C,2013PhRvD..87h5037C,2018JCAP...11..019C,2020JCAP...06..011Z}. FFC can also be triggered by various mechanisms with the interplay between neutrino transport and matter interactions in CCSN (see, e.g., \cite{2019PhRvL.122i1101C,2020PhRvR...2a2046M,2021PhRvD.104h3025N,2021PhRvD.104f3014N,2022JCAP...03..051A}).

In the last few years, it has been demonstrated that the neutrino-matter interactions, in particular momentum-exchanged scatterings, affect flavor conversion in non-linear phase \cite{2021PhRvD.103f3002S,2021arXiv210914011S,2021PhRvD.103f3001M,2021ApJS..257...55K,2022PhRvD.105d3005S,2022arXiv220411873H}. In these numerical models, we have witnessed that the effect (enhancement or suppression of flavor conversion) depends on not only weak processes but also initial condition (e.g., angular distributions of neutrinos), and numerical setup (homogeneous or inhomogeneous). This exhibits the importance of self-consistent treatment of neutrino transport, collision term, and flavor conversion, i.e., the need of global simulations. This motivates us to incorporate collision term into GRQKNT code.

The collision term is implemented into GRQKNT by following the same approach of \cite{2019PhRvD..99l3014R} (see also \cite{1993PhRvL..70.2363R,1996slfp.book.....R,2000PhRvD..62i3026Y,2014PhRvD..89j5004V,2016PhRvD..94c3009B} for more general discussions in treatments of collision term). In the current version, four major weak-processes relevant to CCSN and BNSM are implemented; we describe the essence below. Incorporating more reactions and including high-order corrections on each weak process are postponed to future work.

\subsection{Emission and absorption}\label{sec:emisabs}

\begin{figure*}
   \includegraphics[width=\linewidth]{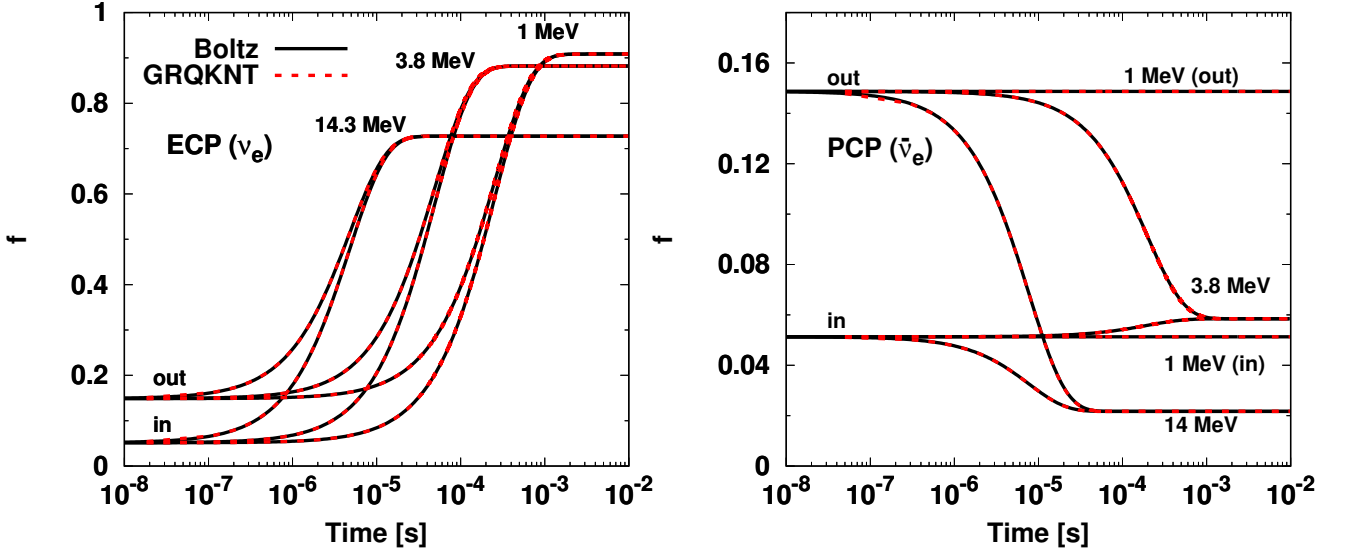}
   \caption{Time evolution of $ee$ component of density matrix in a test of emission and absorption processes. Left: electron capture by free proton (and the inverse process, $\nu_e$ absorption). Right: positron capture by free neutron (and its inverse process, $\bar{\nu}_e$ absorption). We show the result of 1 MeV, 3.8 MeV, and 14.3 MeV neutrinos traveling in outgoing- ($\cos \theta_{\nu}=1$) and incoming ($\cos \theta_{\nu}=-1$) directions. Black solid-lines and red dashed-ones are the results computed by our Boltzmann code and GRQKNT one, respectively.
}
   \label{graph_Emisabstest}
\end{figure*}

\begin{figure*}
   \includegraphics[width=\linewidth]{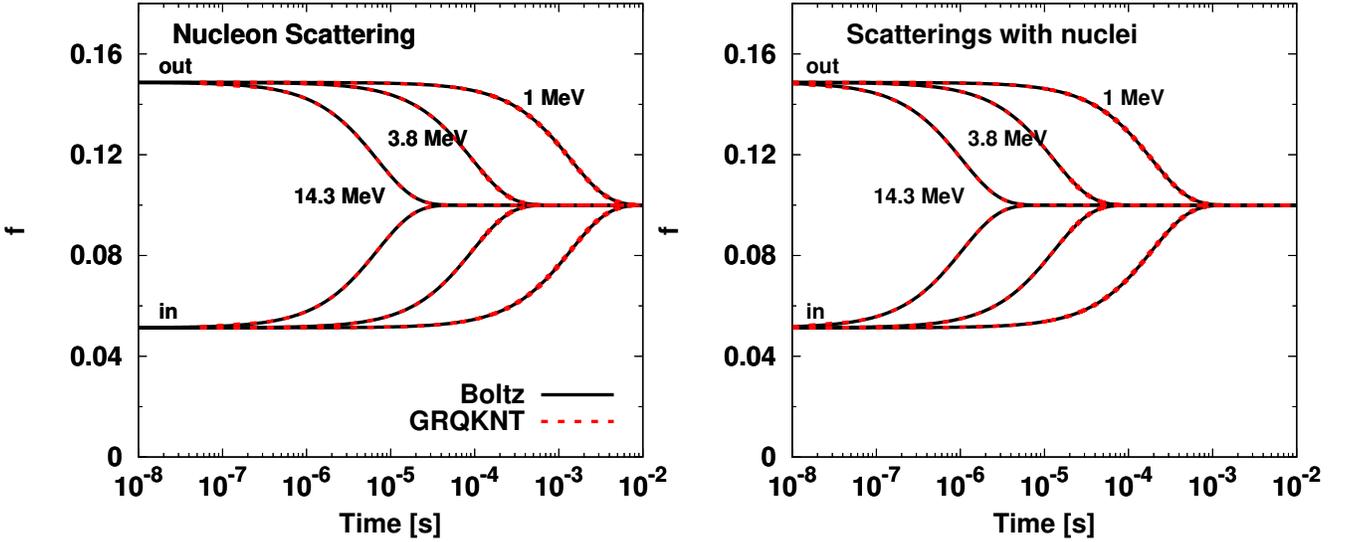}
   \caption{Same as Fig.~\ref{graph_Emisabstest} but for nucleon scatterings (left) and coherent scatterings with heavy nuclei (right). Since there are no flavor dependence in the two weak processes due to neglecting high-order corrections, we only show the result of $f_{ee}$.
}
   \label{graph_Scattest}
\end{figure*}

The two emission processes, electron capture by free protons and positron capture by free neutrons, and their inverse reactions, i.e., neutrino absorptions, are the dominant charged-current reactions in CCSN and BNSM. The neutrino production ($\barparena{R}_{emis}$) and extinction rates ($\barparena{R}_{abs}$) in classical Boltzmann equation can be given as,
\begin{equation}
  \begin{split}
&\barparena{R}_{emis} = \barparena{j}_{e} (1 - \barparena{f}_{ee}), \\
&\barparena{R}_{abs} = \barparena{\kappa}_{e} \barparena{f}_{ee},
  \end{split}
\label{eq:ColClaEmisAbs}
\end{equation}
where $\barparena{j}_{e}$ and $\barparena{\kappa}_{e}$ denote the emissivity and absorption opacity, respectively. Given neutrino energy and the chemical composition of electron (positron), proton, and neutron\footnote{To obtain these thermodynamical quantities, we need to specify an equation-of-state (EOS). In the current version of GRQKNT, we employ a nuclear EOS of \cite{2017JPhG...44i4001F}, which has been used in our CCSN simulations with full Boltzmann neutrino transport (see, e.g., \cite{2019ApJS..240...38N,2019ApJ...880L..28N}).}, $\barparena{\kappa}_{e}$ can be computed from $\barparena{j}_{e}$ with a detailed-balance relation. The emissivity and absorption opacity are computed based on \cite{1985ApJS...58..771B}, which ignores high-order corrections (such as recoil effects) but captures the essential properties of these reactions.

As shown in \cite{2019PhRvD..99l3014R}, those emission and absorption processes of charged-current reactions can be extended in quantum kinetic treatments. It can be written as,
\begin{equation}
\barparena{S}_{ab} = 
  \barparena{j}_{a} \delta_{ab}
- \biggl(   \langle \barparena{j} \rangle_{ab}  + \langle \barparena{\kappa} \rangle_{ab} \biggr) \barparena{f}_{ab},
\label{eq:ColQKEEmisAbs}
\end{equation}
where the bracket is defined as
\begin{equation}
\langle A \rangle_{ab} \equiv \frac{A_a + A_b}{2}. 
\label{eq:defbracket}
\end{equation}
In the above expression, the indices ($a$ and $b$) specify a neutrino flavor. Since we ignore charged-current processes for heavy leptonic neutrinos, we can set $\barparena{j}_{\mu}=\barparena{j}_{\tau}=\barparena{\kappa}_{\mu}=\barparena{\kappa}_{\tau}=0$.

The computations of these charged-current reactions are straightforward and computationally cheap due to no integral operations in momentum space. To check the correct numerical implementation, we perform a comparison study to our Boltzmann code. In this test, transport- and oscillation operators are switched off. To determine reaction rates, we assume a matter state as $\rho=2 \times 10^{12} {\rm g/cm^3}$, $Y_e=0.3$, and $T=10 {\rm MeV}$, where $\rho$, $Y_e$, and $T$ denote baryon mass density, electron-fraction, and temperature, respectively. As an initial condition of neutrino distributions, we set $\barparena{f}$ as
\begin{equation}
\barparena{f} = \frac{1  +  0.5 \cos \theta_{\nu}}{10} .
\label{eq:fini_coltest}
\end{equation}
It should be mentioned that there are no energy dependence in this initial condition of $\barparena{f}$. Here, we perform the test simulations for three different neutrino energy: $1$ MeV, $3.8,$ MeV and $14.3$ MeV.

The results are displayed in Fig.~\ref{graph_Emisabstest}. Except for $\bar{\nu}_e$ with the energy of 1 MeV in positron capture reaction (see right panel of Fig.~\ref{graph_Emisabstest}), all neutrinos approach the equilibrium state, known as Fermi-Dirac distribution. We also note that the energy-dependent feature of each charged-current reaction is properly captured; for instance, the higher energy of neutrinos settles into the equilibrium state earlier. It should be noted that the emissivity of positron capture process for $1 {\rm MeV}$ $\bar{\nu}_e$ is zero due to the energy threshold of the reaction. Consequently, the $\bar{f}_{ee}$ is constant in time. We confirm that both Boltzmann- and GRQKNT codes give identical results (black solid-lines and red dashed-ones are overlapped each other), ensuring that the emission and absorption terms in GRQKNT are correctly implemented.

\subsection{Scattering}\label{sec:scattering}

We implement two processes of momentum-exchanged scatterings into GRQKNT: nucleon scattering and coherent scatterings with heavy nuclei. We assume that the scatterings are elastic, which are reasonable assumptions in CCSN and BNSM. The resultant collision term has a similar form as that of classical Boltzmann equation, that can be written as,
\begin{equation}
  \begin{split}
\barparena{S}_{ab} (\nu^{F},\Omega^{F}) & = - \frac{(\nu^{F})^{2}}{(2 \pi)^3} \int d \Omega^{{\rm \prime F}} R (\nu^{F},\Omega^{F},\Omega^{{\rm \prime F}}) \\
& \times \Bigl( f^{F}_{ab} (\nu^{F},\Omega^{F}) - f^{F}_{ab} (\nu^{F},\Omega^{{\rm \prime F}}) \Bigr),
  \end{split}
\label{eq:ColElaSca}
\end{equation}
where the superscript (F) is put on the variables measured in the fluid-rest frame. For nucleon scattering processes, energy-dependence in the reaction kernel $R$ can be dropped in our assumptions (no recoils and weak-magnetism). We compute these reaction rates by following \cite{1985ApJS...58..771B}.

We perform a similar test-simulation as that used in emission and absorption processes. We employ the same matter background (to determine the reaction rate) and initial neutrino distributions as those used in Sec.~\ref{sec:emisabs}. In this test, we focus only on $\nu_e$, since the two scattering processes do not depend on neutrino species in our assumptions.

The results are summarized in Fig.~\ref{graph_Scattest}. As expected, $f_{ee}$ evolves towards isotropic distributions. It should be noted that the initial angular distribution does not depend on neutrino energy; hence, the isotropic distribution becomes the same among different energy of neutrinos, which is why all lines in the figure converges to the same value. On the other hand, the time-evolution of $f_{ee}$ is energy dependent, in which the higher energy of neutrinos achieve isotropic distributions earlier. We confirm that the result of GRQKNT is good agreement with Boltzmann simulation, illustrating the correct implementation.

As another test related to collision term, we perform a homogeneous simulation of FFC with scatterings. This corresponds to a representative example to assess the capability of GRQKNT for problems coupling neutrino oscillations with scatterings; we shall present the result in the next section.

\section{Oscillation module}\label{sec:osc}
Implementing neutrino oscillation module is the most important upgraded element from our classical Boltzmann solver. Aside from a requirement of high-order accuracy of time integration scheme, the numerical treatment of neutrino oscillation is straightforward. All we need to do is the matrix calculation of $[\barparena{H},\barparena{f}]$, indicating that no numerical instabilities occur\footnote{We note, however, that coarse resolutions in momentum space may generate spurious modes (see, e.g., \cite{2012PhRvD..86l5020S}). This issue should be kept in mind for any numerical simulations of collective neutrino oscillations.}. In this section, we only highlight some representative tests. Most of them are the same as those performed in our previous paper \cite{2021ApJS..257...55K}. We measure the capability of our GRQKNT by comparing to analytic solutions or reproducing the results obtained by previous studies. We also perform inhomogeneous simulations of FFC, in which both neutrino transport and flavor conversion are taken into account. This test shows the applicability of GRQKNT to local simulations, which was discussed in Sec.~\ref{sec:basiceq}. We note that the purpose of this paper is to present the capability of GRQKNT code, and therefore we do not enter into details of physical aspects of each flavor conversion. We refer readers to other references for physics-based discussions of each test.

\subsection{Vacuum Oscillation}\label{sec:vacosc}

\begin{figure*}
   \includegraphics[width=\linewidth]{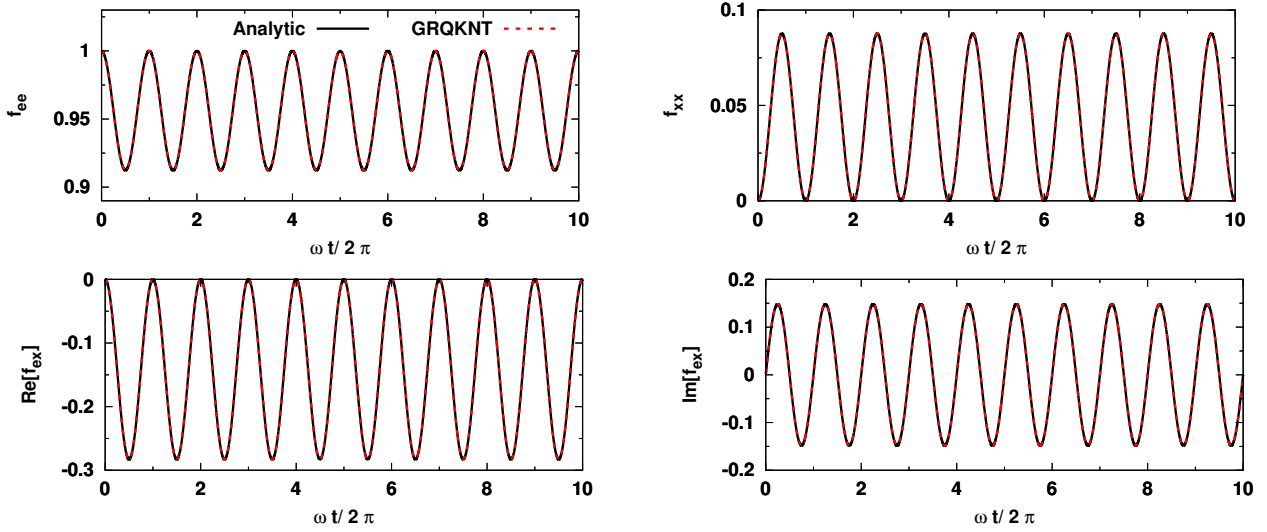}
   \caption{Time evolution of each flavor component of density matrix for vacuum oscillation. Black solid-lines denote the analytic solution, meanwhile the red dashed-ones are the result obtained from GRQKNT simulation. The time is normalized by a vacuum frequency $\omega$, which is defined as ($\omega \equiv \Delta m^2/2 \nu$).
}
   \label{graph_VacOSC}
\end{figure*}

We start with a check of vacuum oscillation. In this test, we assume normal ordering of neutrino masses. The oscillation parameters are set as $\Delta m^2 = 2.45 \times 10^{-15} {\rm MeV^2}$ and $\sin^2 \theta_0 = 2.24 \times 10^{-2}$ where $\Delta m^2$ and $\theta_0$ denotes the neutrino squared-mass difference and mixing angle under two-flavor approximation, respectively. The neutrino energy is assumed to be $20$ MeV. Our parameter choice is the same as that used in \cite{2021ApJS..257...55K} (see Sec. 5 of the the paper). Fig.~\ref{graph_VacOSC} portrays the time evolution of each component of density matrix of neutrinos. As a reference, we also show the analytic solution in each panel (see Eqs.~14-17 in \cite{2021ApJS..257...55K}). As shown in the figure, our results are good agreement with them.

\subsection{MSW resonance}\label{sec:MSWreso}

\begin{figure*}
   \includegraphics[width=\linewidth]{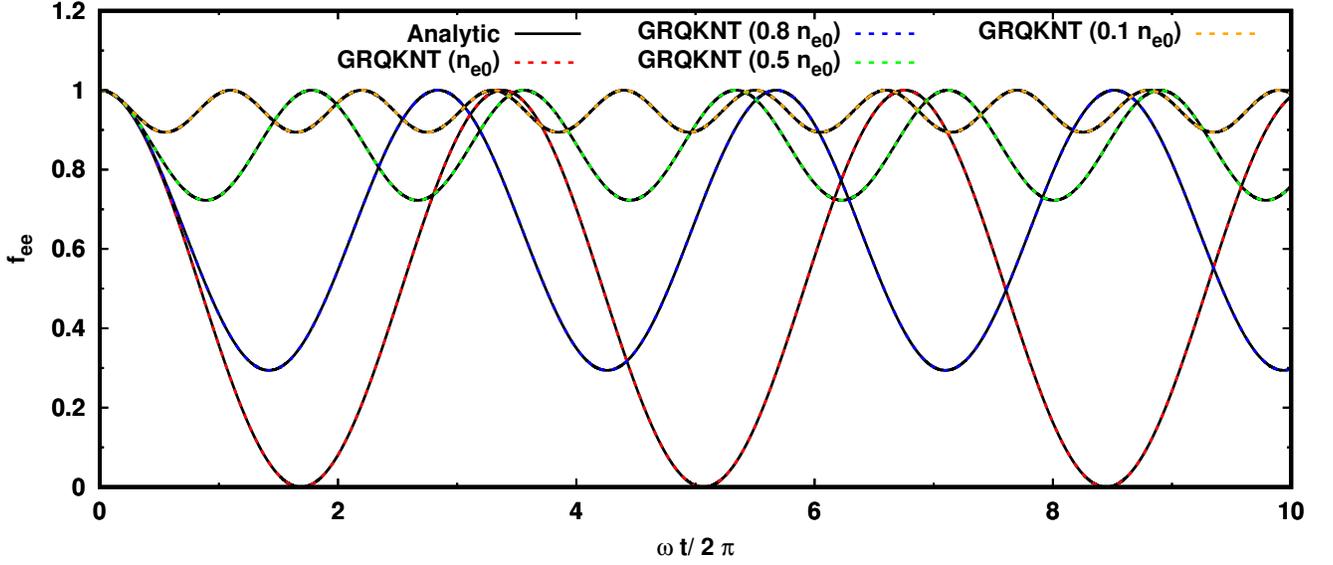}
   \caption{Time evolution of $f_{ee}$ for MSW neutrino oscillation. Black solid-lines denote the analytic solutions. Color distinguish our simulation models with varying the number density of electron, where $n_{e0}$ denotes the resonance density (see Eq.~\ref{eq:ne0}).
}
   \label{graph_MSW}
\end{figure*}

To check implementation of matter Hamiltonian, we carry out a simulation of MSW effect. Assuming homogeneous electron distributions, the solution can be derived analytically (see Sec. 6 in \cite{2021ApJS..257...55K}). Under the two flavor approximation with normal mass hierarchy, the resonant electron-number density can be written as,
\begin{equation}
n_{e0} = \frac{\Delta m^2 \cos 2 \theta_0}{2 \sqrt{2} G_F \nu}.
\label{eq:ne0}
\end{equation}
With varying electron-number density, 0.1 $n_{e0}$, 0.5 $n_{e0}$, 0.8 $n_{e0}$, and $n_{e0}$, we solved QKE by GRQKNT with the same oscillation parameters as that used in test of vacuum oscillation. As shown in Fig.~\ref{graph_MSW}, the results are good agreement with analytic solutions, exhibiting the correct implementation of matter Hamiltonian.

\subsection{Fast neutrino-flavor conversion (FFC)}\label{sec:FFC}

\begin{figure}
   \includegraphics[width=\linewidth]{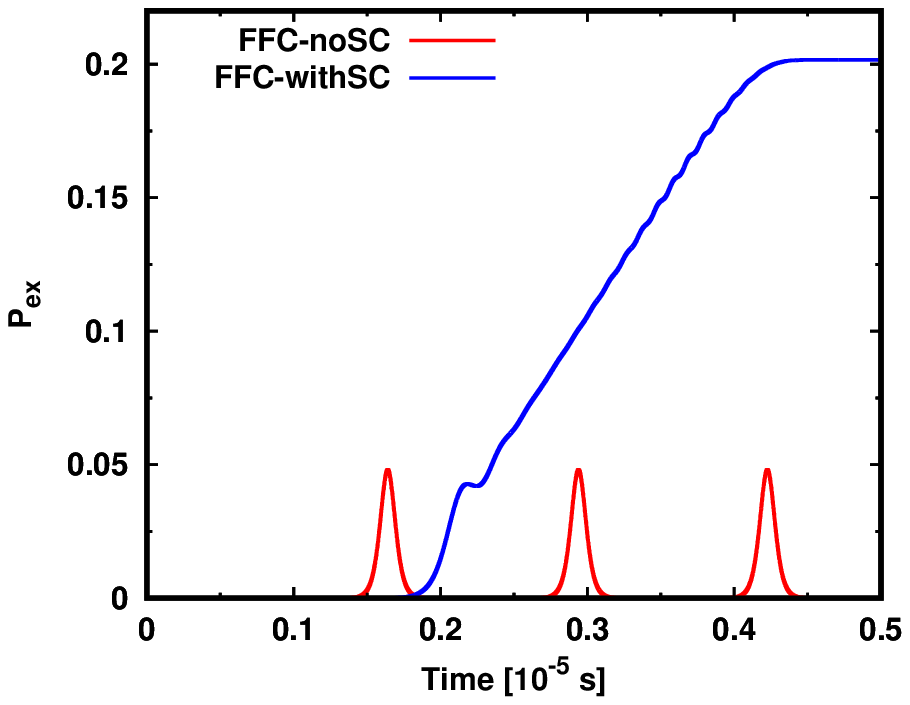}
   \caption{Time evolution of $P_{ex}$ (see Eq.~\ref{eq:defPex}) for FFC simulations. Red (blue) line represents the result of FFC without (with) isoenergetic scatterings.
}
   \label{graph_FFChomo}
\end{figure}

\begin{figure}
   \includegraphics[width=\linewidth]{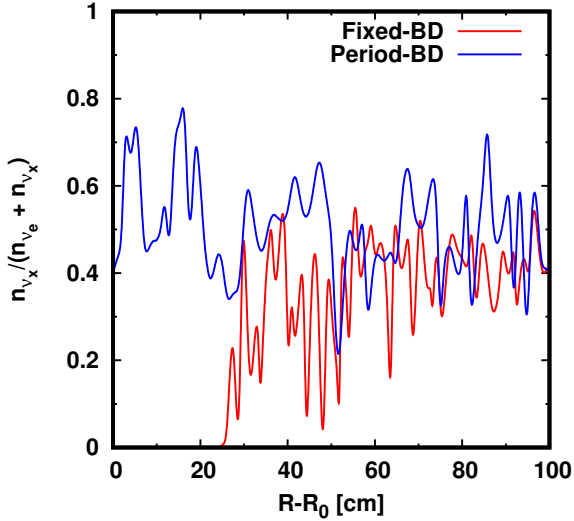}
   \caption{Radial profile of number density of $\nu_x$ ($n_{\nu_x}$) normalized by the flavor-integrated one ($n_{\nu_e} + n_{\nu_x}$) for inhomogeneous FFC simulations. Color distinguishes models: red (fixed-boundary) and blue (periodic-boundary). We display the result at $t=10^{-8}$ s.
}
   \label{graph_FFC_inhomo_radi_vs_numden}
\end{figure}

\begin{figure*}
   \includegraphics[width=\linewidth]{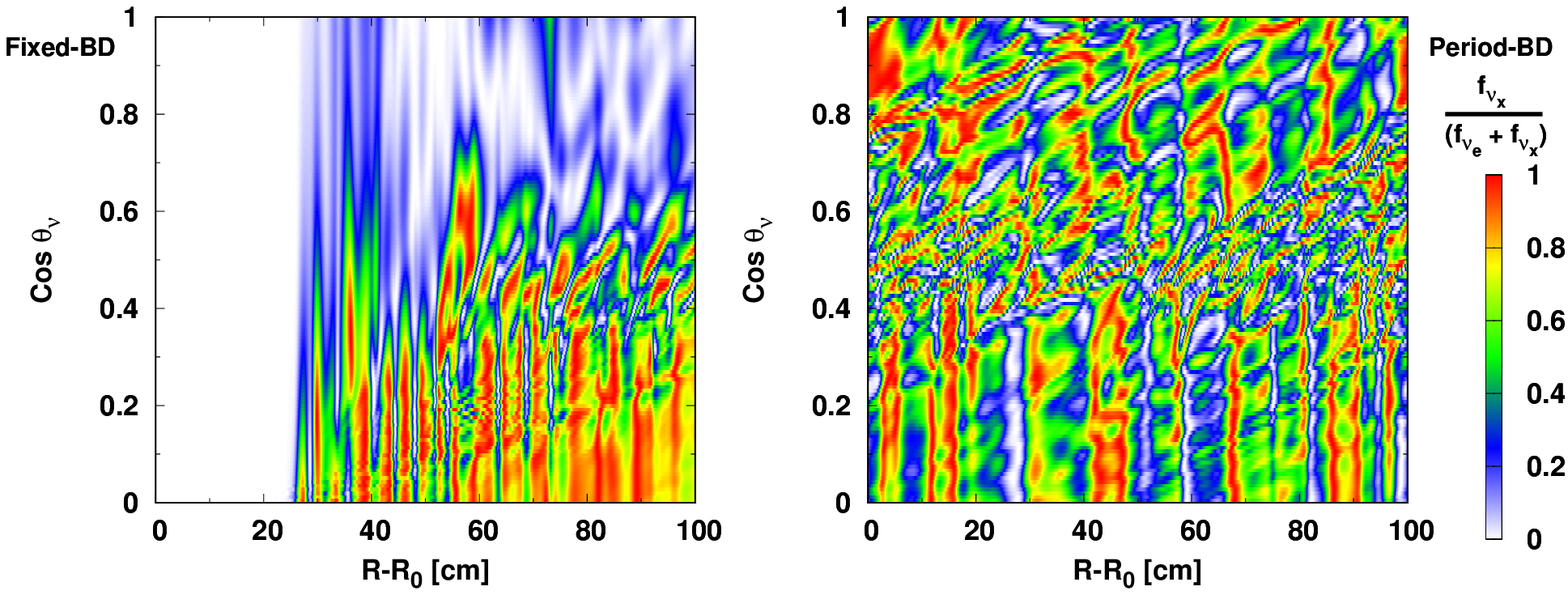}
   \caption{Radial and angular distributions of neutrinos for two local inhomogeneous simulations of FFC. The color map displays $f_{\nu_x}/(f_{\nu_e} + f_{\nu_x})$. $R_0$ is set to be $50$ km, and the computational domain is $\Delta R = 100$ cm. The left panel shows the result of simulation with a fixed boundary, in which the neutrino angular distribution in $\cos \theta_{\nu} > 0$ flight directions at the inner boundary ($R-R_{0} = 0$) is constant in time. The right one shows the result in the case with periodic boundary condition. We display the result at $t=10^{-8}$ s. At that time, the neutrino distributions have already settled into a quasi-steady state.
}
   \label{graph_Radi_vs_angular_inhomo}
\end{figure*}

We assess the capability of GRQKNT to problems that self-interaction potential plays a dominant role in flavor conversion. As a representative example, we adopt fast neutrino-flavor conversion (FFC) in this test. We perform homogeneous multi-angle simulations of FFC with and without iso-energetic scatterings. Since there are no analytic solutions for FFC problems which neutrinos in all different angles of momentum space are coupling, we compare our results to those obtained in previous studies \cite{2021PhRvD.103f3002S,2021ApJS..257...55K}. We also carry out simulations of FFC with transport terms, that exhibits the applicability of GRQKNT to any local inhomogeneous simulations.

The initial condition for homogeneous simulations is set as follows. The angular distributions of $f_{ee}$ and $\bar{f}_{ee}$ are set as (see also \cite{2021PhRvD.103f3002S,2021ApJS..257...55K}),
\begin{equation}
  \begin{split}
& f_{ee} = 0.5 C \\
& \bar{f}_{ee} = \biggl[ 0.47 + 0.05 \exp( -(\cos \theta_{\nu}-1)^2   )   \biggr] C,
  \end{split}
\label{eq:initialFFChomo}
\end{equation}
where $C$ denotes the normalization factor, which is determined so as to be $\mu = 10^5 {\rm km^{-1}}$ ($\mu \equiv \sqrt{2} G_F n_{\nu}$ where $n_{\nu}$ denotes the number density of $\nu_e$). We assume that other components of density matrix are zero. We trigger FFC by adding a vacuum potential, whose parameters are set as $\theta_0 = 10^{-6}$, $\Delta m^2 = 2.5 \times 10^{-6} {\rm eV^2}$. The neutrino energy is assumed to be $50$ MeV. In this simulation, we deploy $128$ angular grid points uniformly with respect to $\cos \theta_{\nu}$. Under the same set of these oscillation parameters and initial conditions, we perform another homogeneous simulation with incorporating iso-energetic scattering. The inverse mean free path of scattering is set as $1 {\rm km^{-1}}$ isotropically.

Figure.~\ref{graph_FFChomo} summarizes the results of the two homogeneous FFC simulations. To measure the degree of flavor conversion, we use $P_{ex}$, that is defined as
\begin{equation}
P_{ex} = 1 - \frac{n_{\nu_e}}{n_{\nu_{e0}}},
\label{eq:defPex}
\end{equation}
where $n_{\nu_e}$ and $n_{\nu_{e0}}$ denote the number density of $\nu_e$ and its initial value, respectively. These results are good agreement with previous studies (\cite{2021PhRvD.103f3002S,2021ApJS..257...55K}, and private communication), ensuring reliability of our module computing neutrino self-interaction potential.

Next, we turn our attention to inhomogeneous simulations. As described in Sec.~\ref{sec:basiceq}, GRQKNT is applicable to local simulations by setting a small spatial box that is located away from the origin of spherical polar coordinate. In this test, we set $R_0 = 50$ km (the distance from the coordinate origin) and $\Delta R = 100$ cm (computational domain) to meet the requirement. We also assume spherically symmetric, flat spacetime, and gray neutrino transport. The number of radial and neutrino angular grid points are $N_r = 3072$ and $N_{\theta_{\nu}}=256$, respectively. It should be mentioned that we obtain essentially the same results in low resolution simulations ($N_r=1536$ and $N_{\theta_{\nu}}=128$), suggesting that the adopted resolutions are high enough to capture the essential features. In this test, we solve QKE with the two-flavor approximation, and we set $\nu_x=\bar{\nu}_x=0$ in the initial condition. We run the simulation up to $t=10^{-8}$ s, which is $\sim 3$ times longer than the light crossing time of simulation box for out-going neutrinos. We observe that the system achieves a quasi-steady state by the end of our simulation.

We set the initial angular distributions of $\nu_e$ and $\bar{\nu}_e$ through a newly-proposed analytic function, which is a simple but has a capability to capture essential characteristics of neutrino angular distributions in CCSN and BNSM. In this model, we focus only on outgoing neutrinos and put negligible atmospheres of neutrinos for incoming neutrinos. We consider a situation that outgoing neutrinos dominate over incoming ones, and that electron neutrinos lepton number (ELN) crossing appears in out-going directions; which would occur in CCSN (e.g., Type-II ELN crossings found in \cite{2021PhRvD.104h3025N}) and BNSM (see, e.g., \cite{2017PhRvD..95j3007W}). The analytic function is written as,
\begin{equation}
f_i = 
\begin{cases}
\langle f_i \rangle \biggl( 1 + \beta_i ( \cos \theta_{\nu} - 0.5 ) \biggr) & \hspace{4mm} \cos \theta_{\nu } \ge 0, \\
\langle f_i \rangle \times \eta & \hspace{4mm} \cos \theta_{\nu } < 0,
\end{cases}
\label{eq:anaAngdistri}
\end{equation}
where the subscript $i$ denotes the neutrino flavor. We set $\eta=10^{-6}$ for negligible contribution of incoming neutrinos to self-interaction potentials. There are two parameters to characterize the angular distribution: $\langle f_i \rangle$ and $\beta_i$. The former is directly associated with the number density of neutrinos, and the latter characterizes the anisotropy of neutrino distributions. Since we need to determine  $\nu_e$ and $\bar{\nu}_e$ angular distributions, we have four parameters in total. Below, we describe how we determine them.

In CCSN environment, it has been suggested that ELN crossing tends to occur at the region where the ratio of number density of $\bar{\nu}_e$ to $\nu_e$ becomes unity \cite{2019PhRvD.100d3004A,2019ApJ...886..139N,2021PhRvD.104h3025N}. Hence, we set $\langle f_{ee} \rangle = \langle \bar{f}_{ee} \rangle$ in this test. We also note that $\bar{\nu}_e$ tends to be more forward-peaked angular distributions than $\nu_e$, i.e., $0<\beta_{\nu_e} < \beta_{\bar{\nu}_e}$, which is mainly due to the disparity of neutrino-matter interactions between $\nu_e$ and $\bar{\nu}_e$. As a simple case, we set $\beta_{\nu_e} = 0$ and $\beta_{\bar{\nu}_e} = 1$ in this test. We note that the ELN crossing is located at $\cos \theta_{\nu} = 0.5$ when we set $\langle f_{ee} \rangle = \langle \bar{f}_{ee} \rangle$ regardless of the choice of $\beta$ for both species. We determine $\langle f_{ee} \rangle$ so that the number density of $\nu_e$ becomes $6 \times 10^{32} {\rm cm^{-3}}$. This number density corresponds to the case that $\nu_e$ has a luminosity of $\sim 4 \times 10^{52} {\rm erg/cm^3}$ with an average energy of $\sim 12 {\rm MeV}$ at $R = 50$~km. These are typical values for each variable at a few hundred milliseconds after bounce in CCSN. The systematic study for the parameter dependence is very interesting, which is, however, postponed to our future work.

We carry out two simulations with different boundary conditions. One of them is to use a fixed boundary condition at $R=R_{0}$, in which the neutrino distributions are frozen. More precisely speaking, the initial value of the density matrix of neutrinos with $\cos \theta_{\nu} > 0$ in the first radial grid point is restored at each time step. On the other hand, we employ a copy boundary for incoming neutrinos ($\cos \theta_{\nu} < 0$). As an outer boundary condition, we set a copy boundary for $\cos \theta_{\nu} > 0$, while we inject dilute neutrino ($\eta \times \langle f_i \rangle$, see Eq.~\ref{eq:anaAngdistri}) in the incoming directions. We start our simulations with setting anisotropic neutrinos (following Eq.~\ref{eq:anaAngdistri}) homogeneously in space.

We find that strong flavor conversion occurs in both cases. Fig.~\ref{graph_FFC_inhomo_radi_vs_numden} displays the radial profile of $n_{\nu_x}/(n_{\nu_e} + n_{\nu_x})$ at the end of our simulation\footnote{This ratio is an appropriate quantity to measure the degree of flavor conversion in our models. Since there are $\nu_x$ are not injected in the simulations, $n_{\nu_x}$ becomes zero if no flavor conversions occur.}. It should be mentioned that the weak flavor conversion in the region of $0~{\rm cm} \le R-R_{0} \lesssim 20~{\rm cm}$ appeared in the fixed boundary simulation (red line in the figure) is an expected result, since the neutrino conversion is artificially suppressed at $R=R_{0}$. Except for the region, the flavor states nearly reaches the equipartition ($n_{\nu_x}/(n_{\nu_e} + n_{\nu_x}) = 0.5$) in both models.

On the other hand, there is a distinct feature in angular distributions of neutrinos between the two models, which can be seen in Fig~\ref{graph_Radi_vs_angular_inhomo}. We measure the degree of flavor conversion by using a variable of $f_{\nu_x}/(f_{\nu_e} + f_{\nu_x})$ in this figure. In the case of fixed-boundary simulation, the flavor conversion is not outstanding for $\cos \theta_{\nu} \gtrsim 0.8$ neutrinos (see left panel), whereas strong flavor conversion emerges regardless of neutrino flight directions in the case with periodic boundary condition (right panel). This result suggests that the boundary condition affects non-linear evolution of FFC. We postpone the detailed analysis how the boundary condition gives an impact on angular distributions of FFC to future work, since this study requires a systematic study with varying neutrino angular distributions and changing the computational domain, that is clearly out of the scope of this code paper.

\section{Summary}\label{sec:summary}
In this paper, we describe the design and implementation of our new neutrino transport code GRQKNT with minimum but essential tests. This corresponds to an upgraded solver of full Boltzmann neutrino transport; indeed, we transplanted several modules of our Boltzmann solver to GRQKNT (e.g., two-energy-grid technique). Below, we briefly summarize the capability.

\begin{enumerate}
\item GRQKNT code is capable of solving the time-dependent QKE in the full phase space (six dimension). The transport operator is written in a conservative form of general relativistic QKE (see Sec.~\ref{sec:basiceq}). In the current version, neutrino transport in three different spacetimes (flat spacetime, Schwarzschild black hole, and Kerr black hole) are implemented. The two-energy-grid technique is equipped to treat fluid-velocity dependence self-consistently (see Sec.~\ref{sec:transport}).
\item Major weak processes (neutrino emission, absorption, and scatterings) contributing in collision term are implemented in GRQKNT: electron-capture by free proton (and its inverse reaction, $\nu_e$ absorption), positron-capture by free neutrons (and its inverse reaction, $\bar{\nu}_e$ absorption), nucleon scattering, and coherent scattering with heavy nuclei. Collision term for the flavor off-diagonal components are also taken into account by following \cite{2019PhRvD..99l3014R} (see Sec.~\ref{sec:Colterm}). 
\item Vacuum, matter, and self-interaction Hamiltonian are implemented. The tests demonstrated in Sec.~\ref{sec:osc} lends confidence to our numerical treatment of these oscillation modules.
\end{enumerate}

The versatile design of GRQKNT allows us to study many features of neutrino kinetics, and therefore it will contribute to fill the gap between astrophysics community and neutrino oscillation one. As the first demonstration, we preform time-dependent global simulations of FFC by using GRQKNT, which is reported in another paper \cite{2022arXiv220604097N}. This is an important step to understand astrophysical consequences of FFC in CCSN and BNSM, and we will extend the work to more realistic situation in future.

It should be mentioned, however, that there still remain work needed in developments of GRQKNT. Improving input physics such as neutrino-matter interactions is necessary to study more detailed features of neutrino quantum kinetics. Another shortcoming in GRQKNT code is that it is only applicable to problems with frozen matter background, and the feedback on matter dynamics is completely neglected. This indicates that the radiation-hydrodynamic features with quantum kinetic neutrino transport can not be investigated by the current version of GRQKNT. Although the numerical technique to link to hydrodynamic solver has been already established as demonstrated in our full Boltzmann neutrino transport code, the huge disparity of length and time scales between neutrino flavor conversion and other input physics is a major obstacle. We intend to overcome the issue by involving sub-grid models or developing better methods and approximations in future, although technical and algorithmic innovations are indispensable to achieve this goal. Nevertheless, the present study does mark an important advance towards the first-principle numerical modeling of CCSN and BNSM. We will tackle many unresolved issues surrounding the neutrino quantum kinetics with this code.

\section{Acknowledgments}
H.N is grateful to Chinami Kato, Lucas Johns, Sherwood Richers, George Fuller, Taiki Morinaga, Masamichi Zaizen, and Shoichi Yamada for useful comments and discussions. This research used the high-performance computing resources of "Flow" at Nagoya University ICTS through the HPCI System Research Project (Project ID: 210050, 210051, 210164, 220173, 220047), and XC50 of CfCA at the National Astronomical Observatory of Japan (NAOJ). For providing high performance computing resources, Computing Research Center, KEK, and JLDG on SINET of NII are acknowledged. This work is supported by the Particle, Nuclear and Astro Physics Simulation Program (Nos. 2022-003) of Institute of Particle and Nuclear Studies, High Energy Accelerator Research Organization (KEK).
\bibliography{bibfile}

\end{document}